\documentclass[preprint,12pt]{elsarticle}

\usepackage{amssymb}

\usepackage{inputenc} 
\usepackage{natbib}
\usepackage{subfigure}
\usepackage{graphicx}
\usepackage{amsmath}

\journal{}

\begin{document}
 
\begin{frontmatter}

\title{Identifying the fragment structure of the organic compounds by deeply learning the original NMR data}

\author[1]{Chongcan Li}
\author[2]{Yong Cong}
\author[1]{Weihua Deng\corref{cor1}}
\ead{dengwh@lzu.edu.cn} 
\cortext[cor1]{Corresponding Author} 
\address[1]{School of Mathematics and Statistics, Lanzhou University, Lanzhou 730000, P.R. China} 

\address[2]{College of Chemistry and Chemical Engineering, State Key Lab Appl Organ Chem, Key Lab Nonferrous Met Chem and Resources Utilizat, Lanzhou University, Lanzhou 730000, P.R. China}

\begin{abstract}
We preprocess the raw NMR spectrum and extract key characteristic features by using two different methodologies, called equidistant sampling and peak sampling for subsequent substructure pattern recognition; meanwhile may provide the alternative strategy to address the imbalance issue of the NMR dataset frequently encountered in dataset collection of statistical modeling and establish two conventional SVM and KNN models to assess the capability of two feature selection, respectively. Our results in this study show that the models using the selected features of peak sampling outperform the ones using the other. Then we build the Recurrent Neural Network (RNN) model trained by Data B collected from peak sampling. Furthermore, we illustrate the easier optimization of hyper parameters and the better generalization ability of the RNN deep learning model by comparison with traditional machine learning SVM and KNN models in detail.

\end{abstract}


\end{frontmatter}


\section{Introduction}
\label{}

In the last decades, with the development of algorithm optimization and the improvement of computer hardware, Artificial Intelligence (AI) has made great progress and promotes a new generation of scientific and technological revolution. In particular, the investigation of the deep learning (DL) subjects, accompanied by the advent of the mega data era, has triggered a new wave of artificial intelligence research boom. Today, DL may play the great utilitarian role in automatically processing and analyzing the experimental data of chemical science, e.g., reconstructing and denoising chemical spectra \cite{Ying2017,Klukowski2018,Lee2019}, chemical information calculation and recognition of chemical structures from chemical spectral databases \cite{Meiler2003}, etc.

In the field of organic synthesis, one of the most significant works is to elucidate the chemical structures and measure their composition of the mixture after chemical reaction. Generally, it needs the separation and extraction procedure to isolate each pure compound, and then verify the chemical structure by the intimations provided by chemical spectral information of this pure compound. However, this process of purification is very time-consuming, error-prone,  and has become the biggest bottleneck in chemical synthesis. Chemical spectroscopy, especially Nuclear Magnetic Resonance (NMR) spectrum provides full and accurate feature patterns closely correlated with the structure, dynamics, reaction rate, and chemical environment of molecules. Since NMR phenomenon was initially observed in water and in paraffin by Felix Bloch and Edward Purcell  respectively in 1946, its theory has been undergoing unparalleled development, which made high-resolution liquid NMR spectroscopy, a powerful tool of increasing use and importance in chemistry, pharmaceutical, and biology though its inception in physicist's laboratories. It is the modern analytical spectroscopy techniques that may provide the firmer foundations to facilitate the invention of modern chemical and biological science. If one can determine whether the reaction mixture contains certain molecular structure(s) or not by using DL models learned from the NMR raw data with high-credibility, it will drastically reduce the cost of purification and speed up the synthesis.

In this study, we attempt to do some basic exploration of DL technique for identifying the fragment or substructure of organic compound. The next section describes preprocess of original NMR data and feature selection used in our study. In the third section, the Recurrent Neural Network (RNN) model trained by Data B collected from peak sampling, called the NMRclass, is built and evaluated. Finally, we conclude the paper with some discussions.

\section{Data description and preprocessing}

The NMRglue module in Anaconda python and NMR data processing commercial software MestReNova are employed to perform basic raw data processing. Nmrglue provides robust protocols for reading, writing, and extracting the spectral data-points and experimental parameters stored in a variety of common NMR data formats including Bruker, JCAMP-DX, and NMRpipe amongst others. The basic data processing procedure involves four steps, which is zero filling of the original free induction decay(FID) data, fast Fourier transformation, phase correction, and baseline correction. We start from the data collection and collation. According to the principle of nuclear magnetic spectrum, we carry out feature extraction. Then we rebuild the data set from original NMR spectra used in this study.

Based on the bulit NMR data, two different feature extraction methods are used.  In the first method, according to the characteristic that the chemical shift range of the spectrum is generally from -4 ppm to 16 ppm, we sample 2000 points with equidistance, i.e.,$-4+i/100$, for $i = 0,1,\cdots,1999. $ Then we save the intensities of the corresponding picked points. Finally, the one dimensional vector with 2000 elements is obtained from every spectrum. The second method is to sample the raw data getting through the principle of NMR. As we all know, in a NMR spectrum, the important information includes the peak location, that is the chemical shift, the shape of multiplet peak, and the peak intergal area. Based on this perspective, the second data sampling takes the extreme points of NMR spectrum. From the extreme point, one can get the corresponding chemical shift value and peak height. For this method of data sampling, the final collection of each NMR data is a series of two-dimensional vectors. Among them, the peak heights of the two sampling methods are standardized and reduced to the interval of $[0,5]$; see Figure 1, which are called Data A and Data B, respectively, in this study.

Data A is convenient for the pre-processing on account of equidistant sampling. However, it may miss the information of important peaks, because sometimes the peak is not exactly at the selected points. One more thing is that Data A contains a large number of data with values close to 0 because of the high proportion of noisy signals in NMR spectra, which makes Data  A approximately sparse. Here, we call a vector $x$  approximately sparse, if $\vert x \vert_0 = \lvert \{j; \lvert x_j\rvert > \epsilon  \}\rvert \ll n$, for small values $\epsilon$, for example $\epsilon = 10^{-3}$. As for the second method, the number of peaks varies for different spectra, which increases the difficulty of data storage to a certain extent and further modeling.

In this paper, based on Data A and Data B, we build the AI models, and some comparisons are made. We design and implement the different models of Support Vector Machine (SVM) and K-NearestNeighbor (KNN) for different data sets. Then we evaluate the performance of the models and datasets,  which provides the guidance for the data collection and subsequent study. 

\begin{figure}[htbp]
	\centering
	\subfigure[equidistant sampling]{
		\begin{minipage}[t]{0.48\textwidth}
			\centering
			\includegraphics[height=3.5cm,width=7cm]{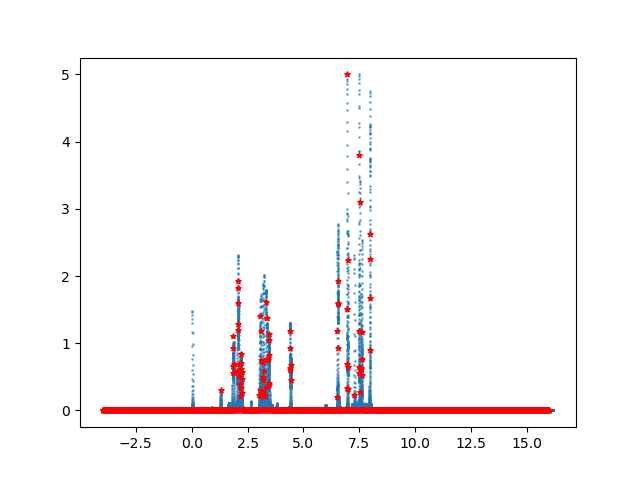}
	\end{minipage}}
	\subfigure[extremum sampling]{
		\begin{minipage}[t]{0.48\textwidth}
			\centering
			\includegraphics[height=3.5cm,width=7cm]{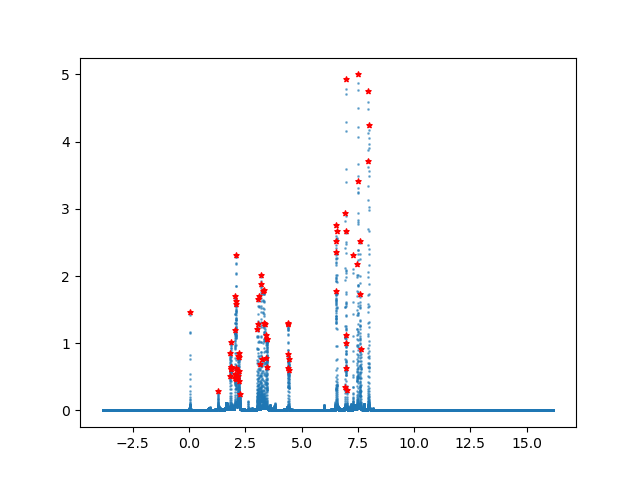}
	\end{minipage}}
	\caption{Feature extraction methods of the original NMR data. Among them, the (blue) data points without mark are the original data points, and the (red) asterisk data points are selected by two different feature extraction methods.}
\end{figure}


\begin{figure}[htbp]
	\centering
	\includegraphics[height=3.5cm,width=8cm]{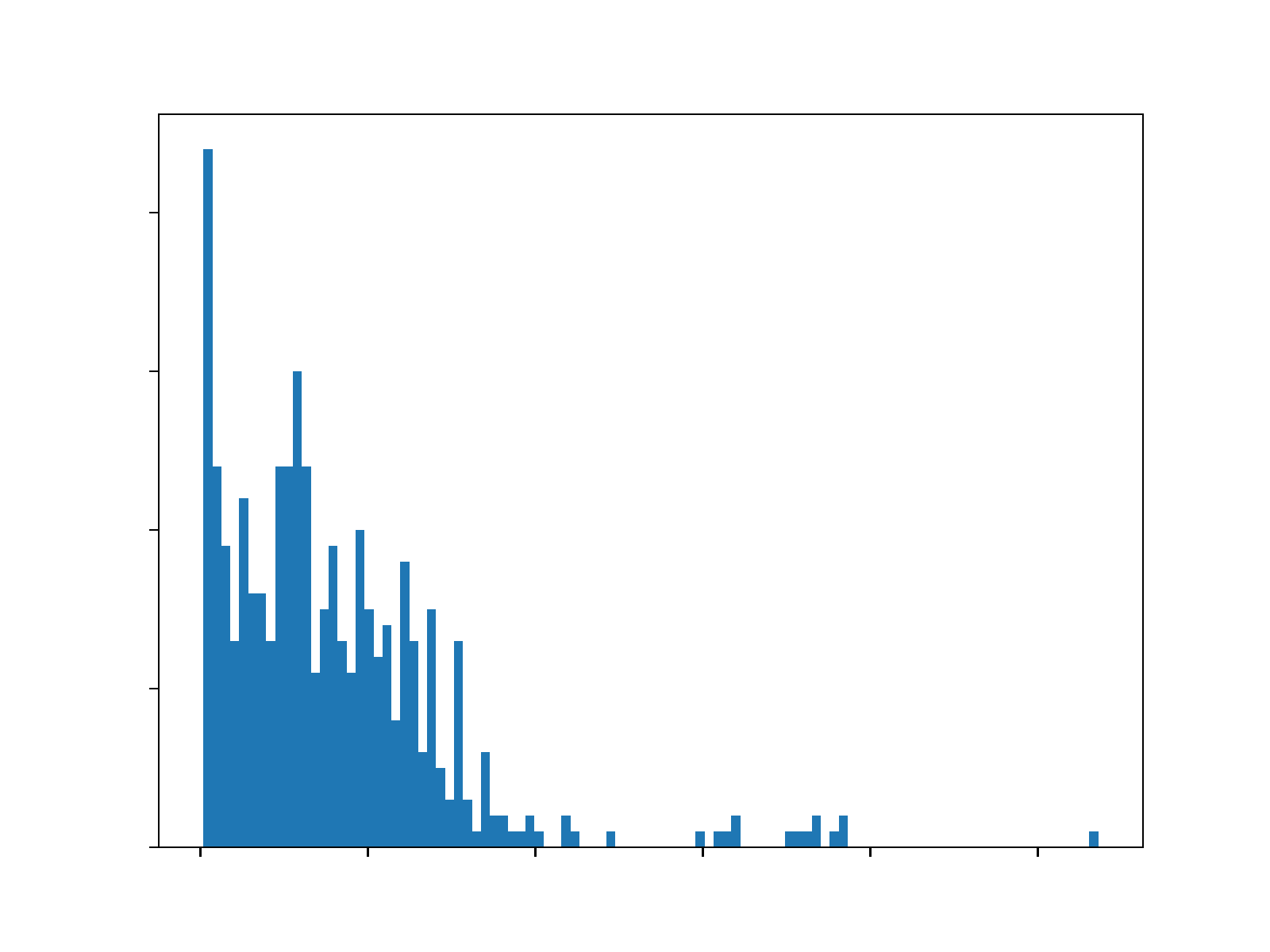}
	\caption{The distribution of character length. Distribution of the number of nonzero elements for Data B. } 
\end{figure}

The above description is for the basic characteristics of the data. Now we take a look at the overall behavior of the data.
In the statistical modeling, there is a simple assumption that the data needs to be evenly distributed. In practice, it is hard to satisfy the requirement of balanced datasets,  which has the adverse impact on the performance of the models.  So we will reveal the imbalance issues and find a way to solve them.

There are several typical data imbalance problems in the original data set of NMR spectra; see Figure 3.
\begin{figure}[htbp]
	\centering
	\includegraphics[height=3.5cm,width=7cm]{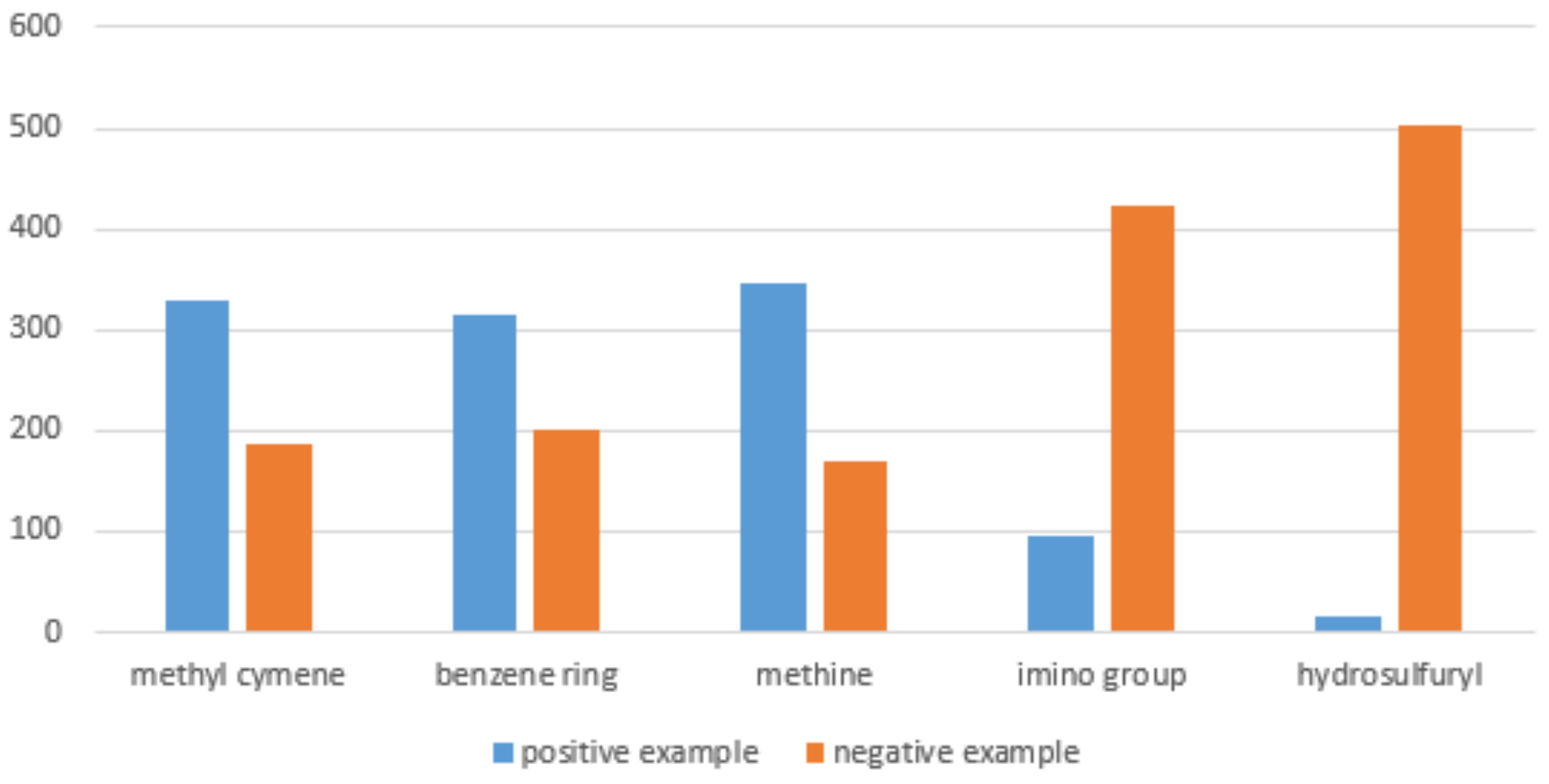}
	\caption{ 
		Data imbalance for some functional groups. Here, the number of positive examples in each kind functional group represents the number of compounds with this structure, and the number of the negative ones is for without this structure. 
	}
\end{figure}

The essence of the impact of unbalanced data  is that positive and negative samples have different ``discourse power" on the model, so the strategy to solve this kind of problem is to increase the weight of a small number of samples. According to the characteristics of the original data of NMR spectra, a new method is proposed to alleviate the imbalance of the original data. We know the series of two dimensional  vector coordinates $(x_i, y_i)$ characterize all information contained in a spectrum. In other words, the recognition of a certain group or structure is based on this set of coordinate points, which is not related to others. Thus, we think that the information of group or molecular fragment contained in the spectrum is not limited by the order of these two-dimensional vector groups. Based on this, in this paper, for the original NMR data, data enhancement methods such as group reversal rearrangement, are carried out to increase the number of samples, alleviate the problem of data imbalance, and improve the performance of the model.

 In order to study how feature selection influences the ability of model recognition. Two models SVM and KNN are established by using  the training sets of Data A and Data B. Four evaluation criteria, accuracy, F1 score, precision, and recall, are constructed by using confusion matrix \cite{Stehman1997}. In this paper, the ratio of the training and testing sample is set as 5:1, and the performance of all models in the test set is shown in Figure 4. The training and prediction are carried out for ten times, respectively, and the best performance result is recorded. Taking the model accuracy as an example, it is between 82.5\% and 91\% for KNN tested in Data A, and it is between 86.5\% and 94.7\% tested in  Data B. Similarly, the accuracy of SVM tested in Data A is between 81.2\% and 91\%, and in Data B is between 84.5\% and 92.5\%. For F1 score, it is between 0.807 and 0.92 for KNN tested in Data A, while for SVM it is between 0.82 and 0.909; for Data B, it is between 0.886 and 0.935 for KNN, while it is between 0.881 and 0.932 for SVM. The overall F1 score result is show in Figure 4 (b).

\begin{figure}[htbp]
	\centering
	\subfigure[(accuracy)]{
		\begin{minipage}[t]{0.48\textwidth}
			\centering
			\includegraphics[height=3.5cm,width=7cm]{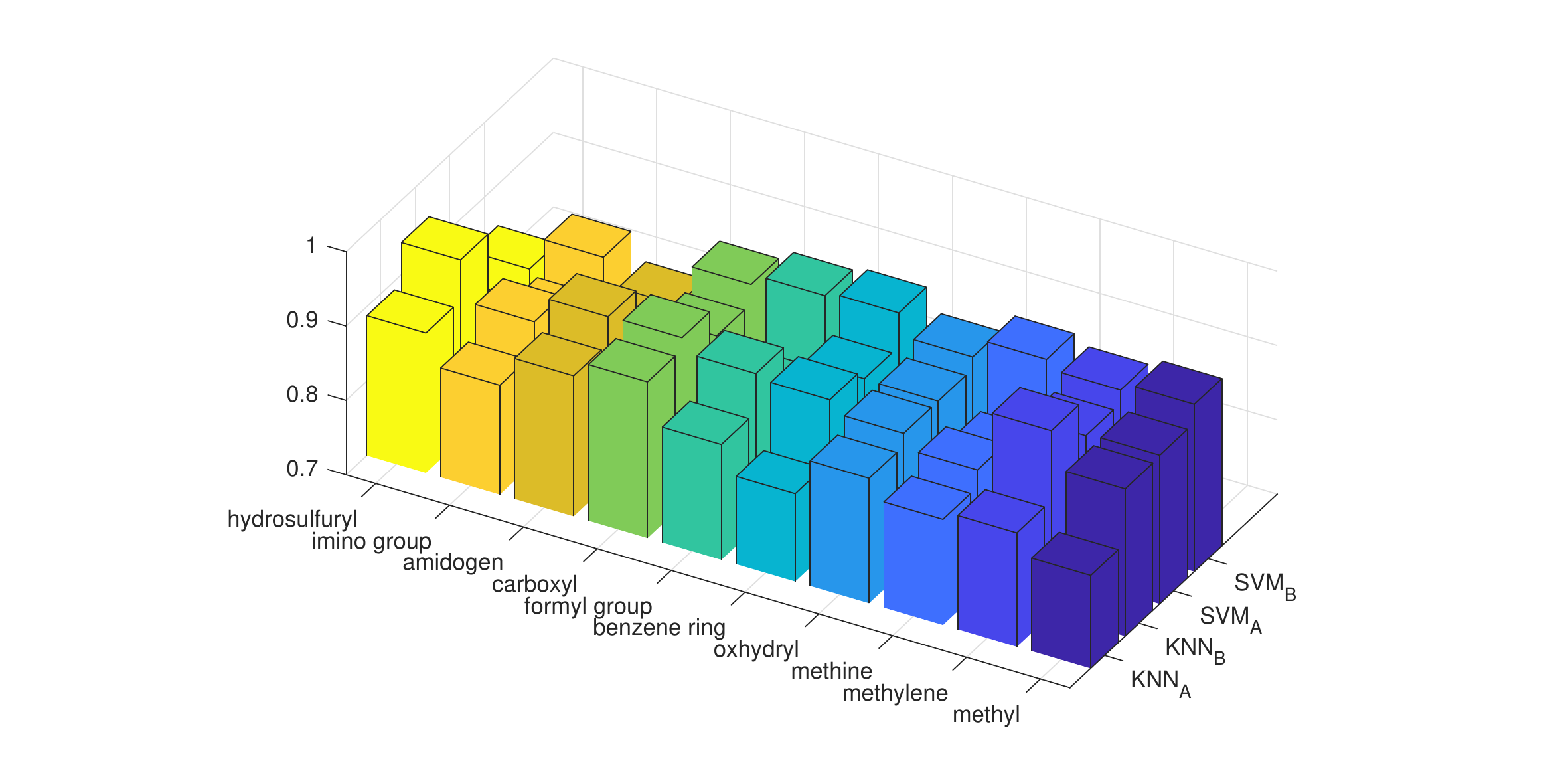}
	\end{minipage}}
	\subfigure[(F1 score)]{
		\begin{minipage}[t]{0.48\textwidth}
			\centering
			\includegraphics[height=3.5cm,width=7cm]{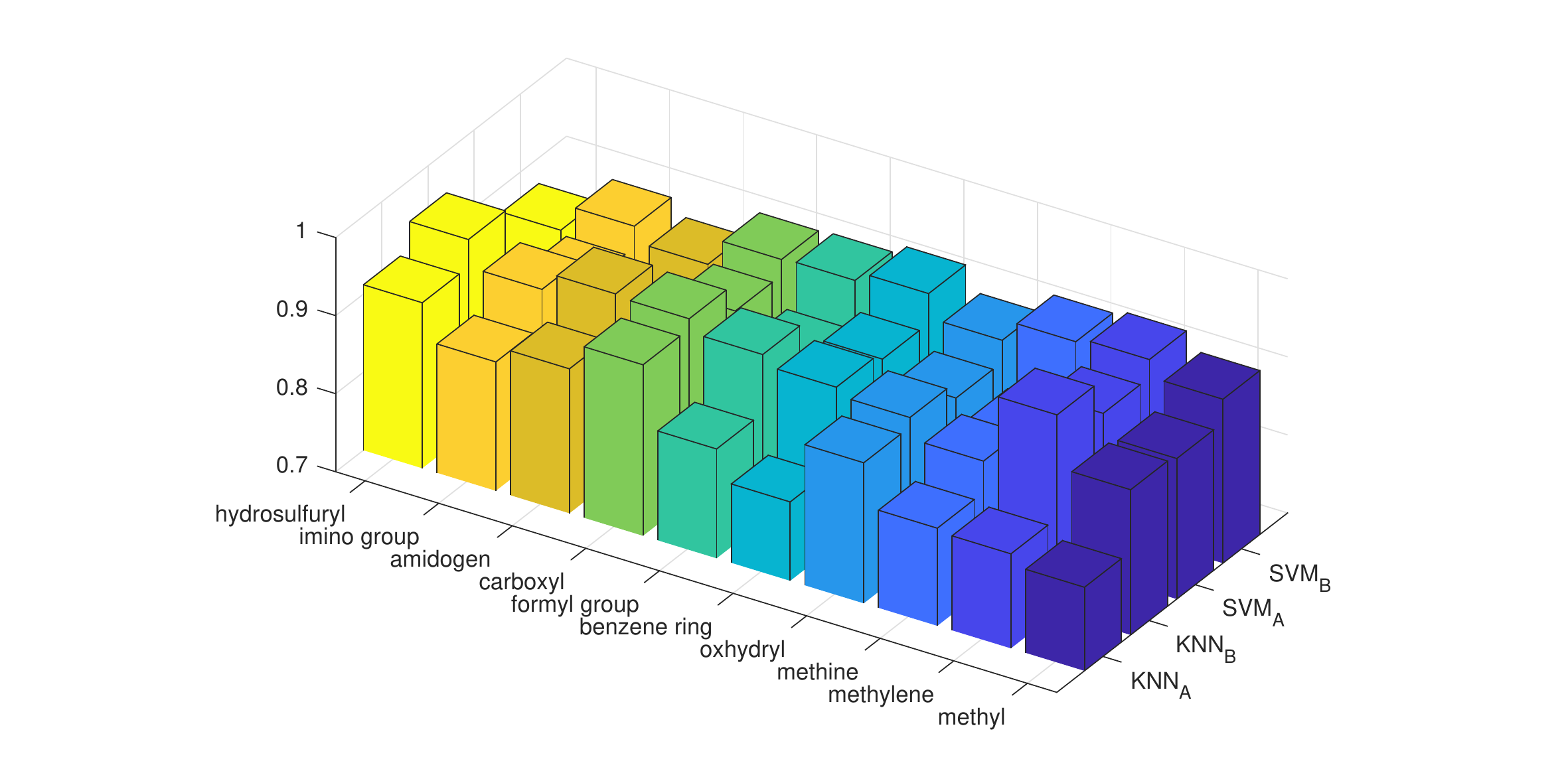}
	\end{minipage}}
	\subfigure[(precision)]{
		\begin{minipage}[t]{0.48\textwidth}
			\centering
			\includegraphics[height=3.5cm,width=7cm]{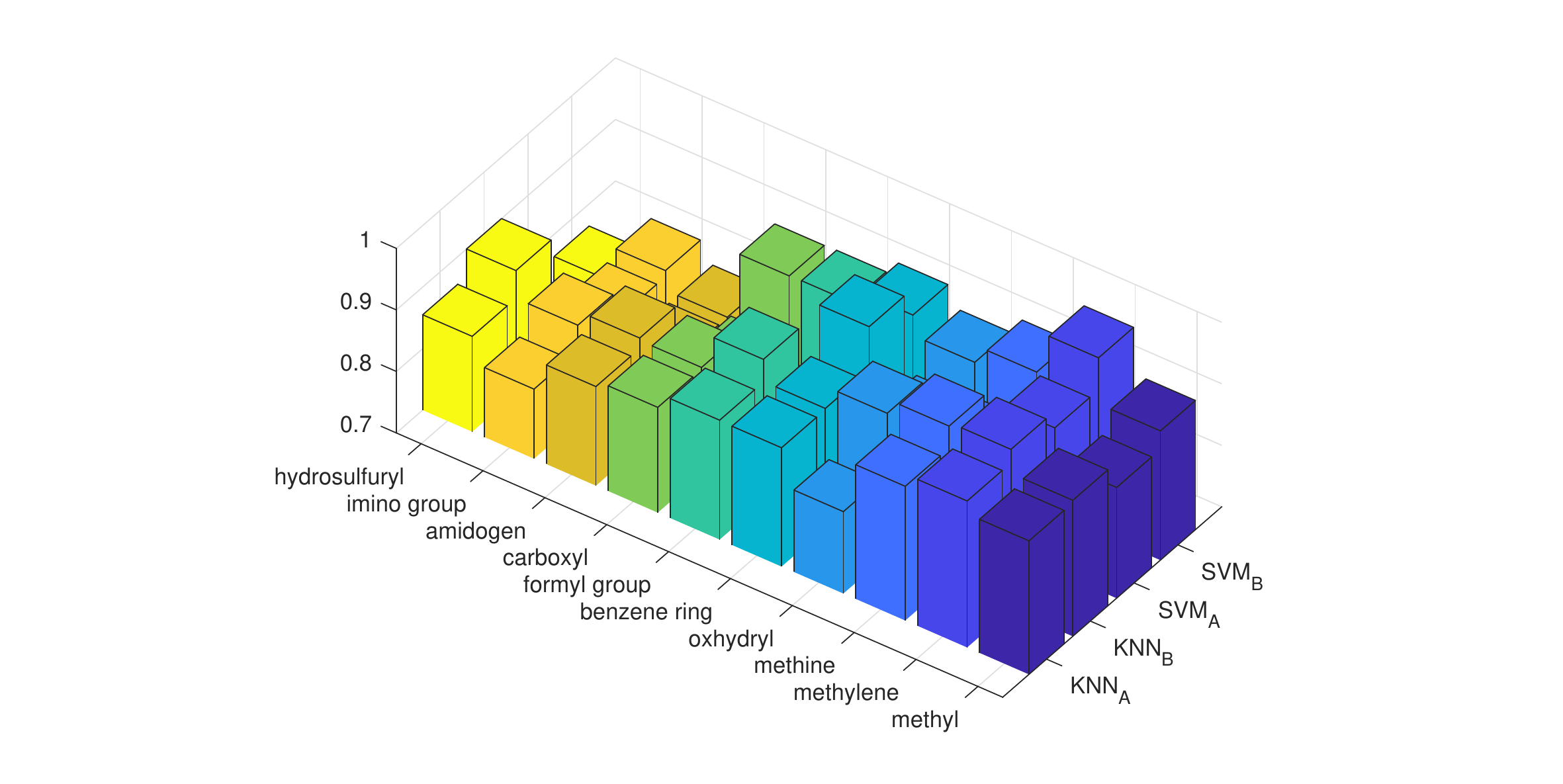}
	\end{minipage}}
	\subfigure[(recall)]{
		\begin{minipage}[t]{0.48\textwidth}
			\centering
			\includegraphics[height=3.5cm,width=7cm]{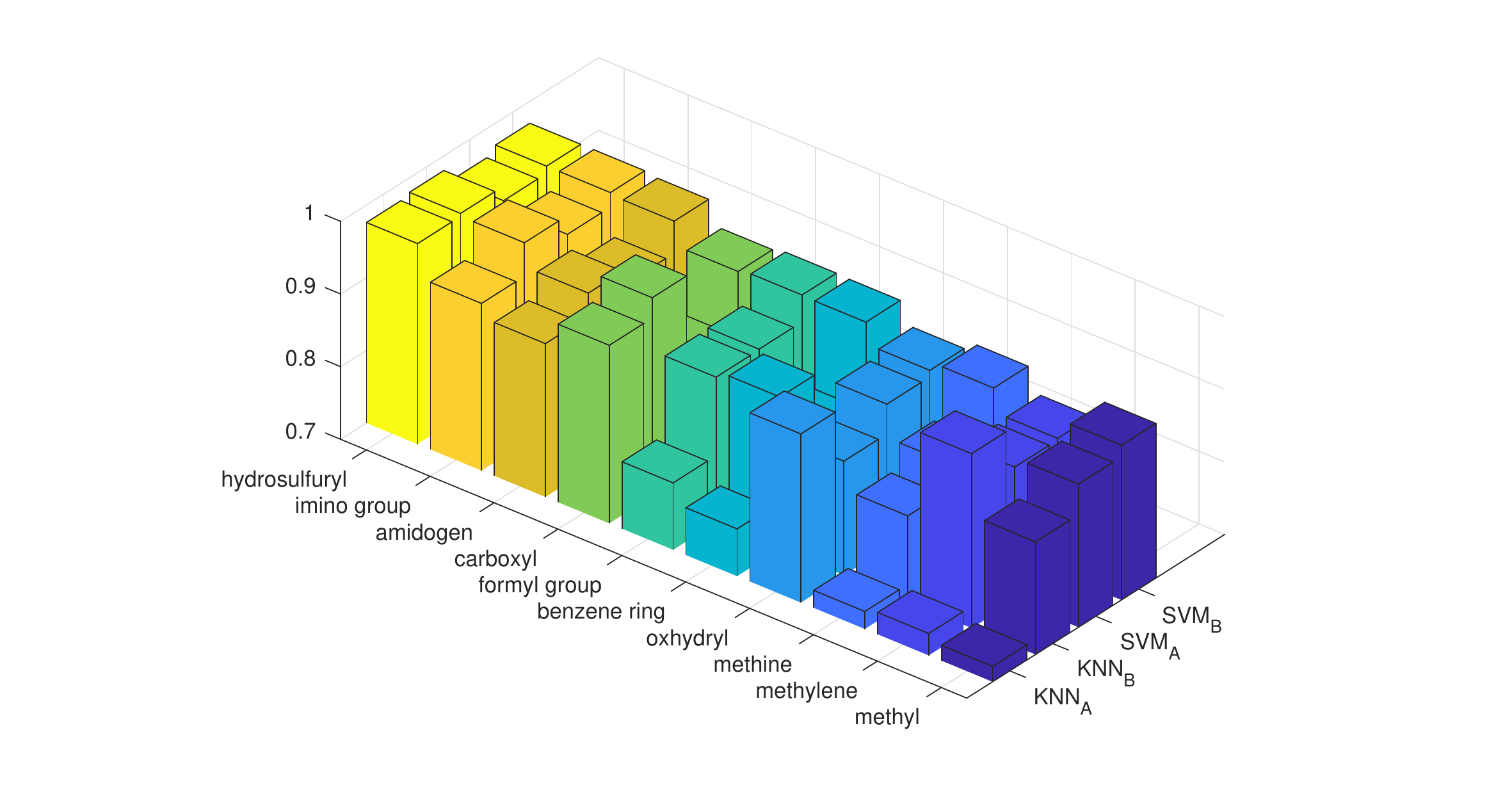}
	\end{minipage}}
	\caption{Performances of SVM and KNN trained on Data A and Data B}
\end{figure}

The above results in Figure 4 show that for each evaluation standard, the height of the KNN$_A$ is always lower than that of the KNN$_B$, and the height of the SVM$_A$ is also lower than that of the SVM$_B$. We know that both Data A and Data B come from the same original data. The models trained on Data B always performs better. Although the collection and processing of Data A is relatively simple, the extraction of key features is still insufficient.  Therefore, the model trained by using the second feature extraction strategy performs better.

\section{Model and experiment}

Based on the insight on the choice of datasets and the experiments of the existing models (SVM and KNN), we build NMRClass, which fundamentally improves the prediction accuracy. Furthermore, we find that free induction decay (fid) NMR data are time sequential signal. Each peak of the corresponding chemical group is correlated with the other ones. In other words, there is a strong correlation between signals in the spectrum. On account of the characteristics of data, we build a deep learning model NMRClass based on recurrent neural network (RNN).

\begin{figure}[htbp]
	\centering	
	\includegraphics[height=4cm,width=8cm]{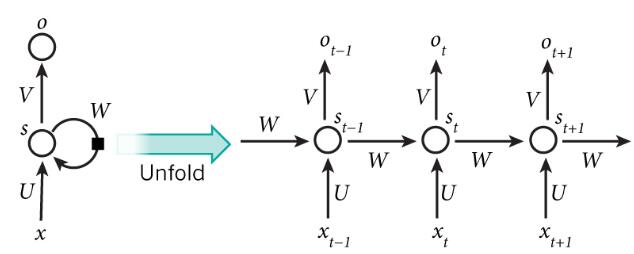}
	\caption{The basic structure of RNN consists of three states: input, hidden, and output, with a cyclic connection. The RNN can be unfolded in time to show the recurrent computation explicitly.}
\end{figure}

First, we introduce the RNN. RNN's output $o_t$ at time t is expressed as follows,
\begin{align*}
	o_t &= \sigma(V\cdot \boldsymbol{s_t}+\boldsymbol{b_{t1}}) \\&= \sigma(V\cdot (U\cdot \boldsymbol{x_t} + W\cdot \boldsymbol{s_{t-1}}+\boldsymbol{b_{t2}})+\boldsymbol{b_{t1}}),
\end{align*}
where $\sigma$ is the activation function, and $U$, $V$and $W$are model parameters \cite{LeCun2015}. With the help of the intermediate variable $\textbf{s}$, the output result at time t is also related to the previous time, which has short-term ``memory ability". Therefore, RNN model well matches the properties of the series data,  and it has achieved good results in text translation, speech analysis,  and other fields \cite{Lipton2015}. RNN is `mnemonic' because it provides a correlation between the previous inputs and the latter ones. The output of the model at the time of $t$ is not only related to the input of $x_t$, but also all previous moments through $\{\textbf{$s_i$}|0 \leq i<t, i \in N \}$.  The chemical shifts and peaks of groups or molecular fragments are  not only  correlated with  the magnetic environment of their own, but also their chemical bonding connections.

In order to accelerate the convergence rate of the model, the activation function selected in the middle hidden layer is ReLU function \cite{Kunihiko1969}, designing for the dichotomous tasks.  And sigmoid function is selected as the activation function of the output layer. Taking the threshold 0.5, if the output is equal to or bigger than 0.5, the output is specified as 1, othervise, it is taken as 0. The traditional SGD algorithm is improved by momentum \cite{Bottou1998}. The hyper parameters selected in Adam are as follows: 
$$\alpha=0.001, \beta_1=0.9, \beta_2=0.999, {\rm and } ~ \epsilon=10^{-4}.$$

To solve the problem of uneven distribution in the data, the  Focal-Loss function is used in this study, i.e.,
$$ Focal-Loss=\left\{
\begin{aligned}
	-\alpha(1-p)^\gamma log(p)  \quad &y=1,\\
	-(1-\alpha)p^\gamma log(1-p)  \quad &y=0.
\end{aligned}
\right.
$$
In fact, the well known cross entropy loss function is just a special case of the above Focal-Loss function \cite{Rubinstein2004}. The hyper paramters are taken as
$$\gamma=2 ~{\rm and } ~\alpha=0.25,$$ which represent
the proportion of positive and negative samples in the data, and play an 
important role in the problem of data imbalance.


The batch data processing is selected for the model with the batch size 16, and the iteration number of model training is set to 300. Aiming at the problem of data imbalance, the method of  data oversampling  described the above section is adopted. In order to prevent over fitting, the model also adopts the dropout technique, with hyper parameter 0.5.

The accuracy of all classification results of deep learning model NMRclass on Data B is shown in Figure 6. The overall performance of deep learning model NMRClass and the comparison of KNN and SVM machine learning model are listed inTable 1.

\begin{figure}[htbp]
	\centering
	\includegraphics[height=3.5cm,width=7cm]{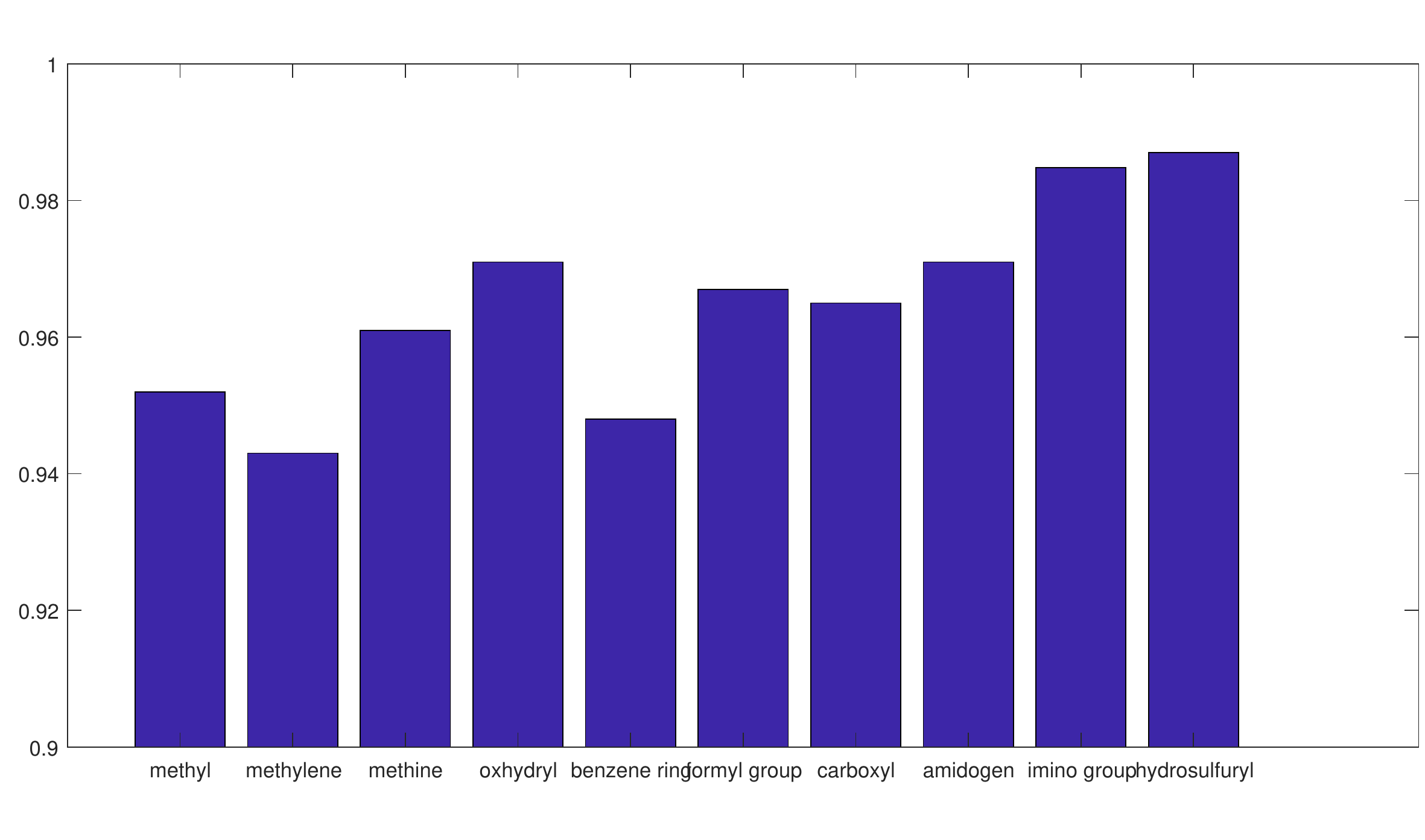}
	\caption{Performance of NMRClass tested on Data B}
\end{figure}

\begin{table}[h]
	\centering
	\begin{tabular}{|l|l|l|l|l|}
		\hline
		& \textbf{macro-acc} & \textbf{marco-precision} & \textbf{macro-recall} & \textbf{marco-f1} \\ \hline
		\textbf{SVM}      & 0.903              & 0.8797                   & 0.9217                & 0.908             \\ \hline
		\textbf{KNN}      & 0.9086             & 0.9063                   & 0.9151                & 0.9139            \\ \hline
		\textbf{NMRClass} & \textbf{0.97238}   & \textbf{0.9432}          & \textbf{0.9736}       & \textbf{0.958}    \\ \hline
	\end{tabular}
	\caption{Performance for the three models }
\end{table}

Figure 6 shows the generalization ability of NMRclass model, that is, the prediction performance for test set, which implies that the structure of recurrent neural network can effectively extract the features of NMR spectra. The success of NMRClass mainly comes from two folds: 1. Data B is built by good sampling; 2. the strong learning ability of deep learning neural network.

To inspire the forthcoming exploration and research, in Figure 7 and Figure 8, we respectively show the training procedure and the structure of NMRClass.

\begin{figure}[htbp]
	\centering
	\subfigure[training-loss]{
		\begin{minipage}[t]{0.48\textwidth}
			\centering
			\includegraphics[height=5.5cm,width=7cm]{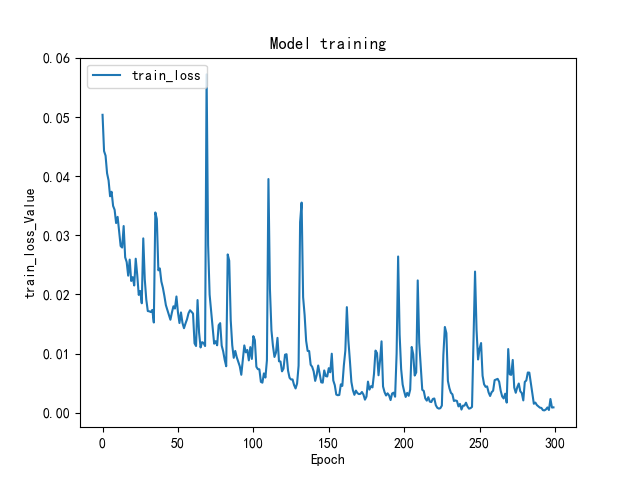}
	\end{minipage}}
	\subfigure[training-accuracy]{
		\begin{minipage}[t]{0.48\textwidth}
			\centering
			\includegraphics[height=5.5cm,width=7cm]{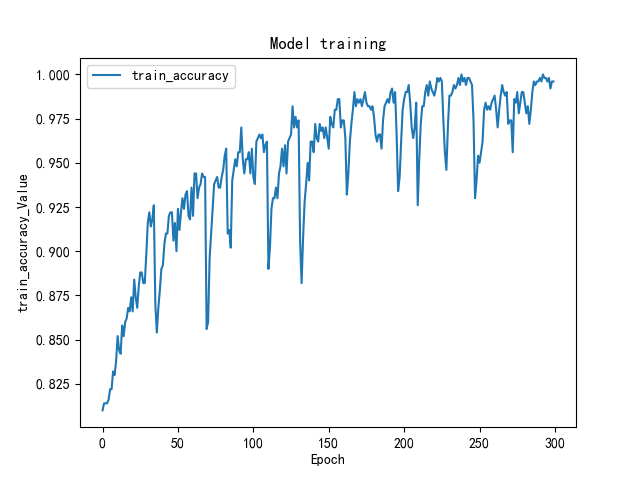}
	\end{minipage}}
	\subfigure[test-loss]{
		\begin{minipage}[t]{0.48\textwidth}
			\centering
			\includegraphics[height=5.5cm,width=7cm]{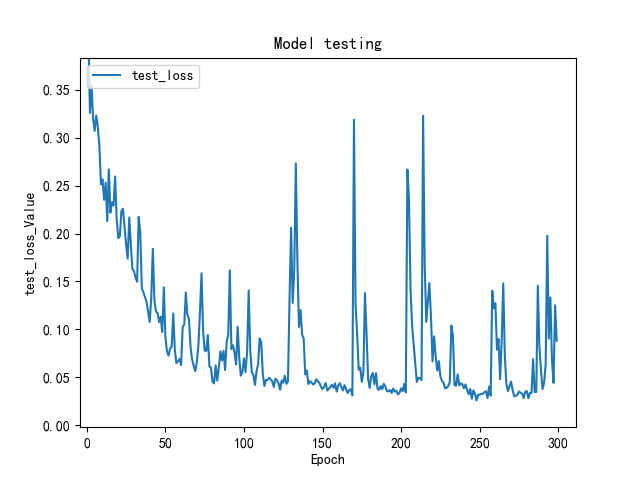}
	\end{minipage}}
	\subfigure[test-accuracy]{
		\begin{minipage}[t]{0.48\textwidth}
			\centering
			\includegraphics[height=5.5cm,width=7cm]{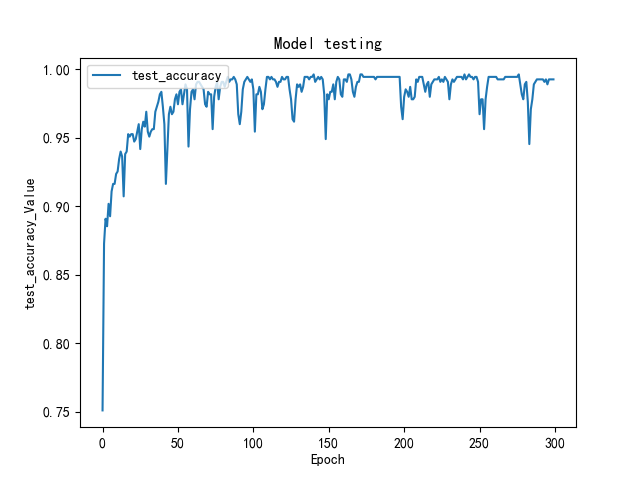}
	\end{minipage}}
	\caption{Training and testing}
\end{figure}

\begin{figure}[htbp]
	\centering
	\includegraphics[height=12cm,width=7cm]{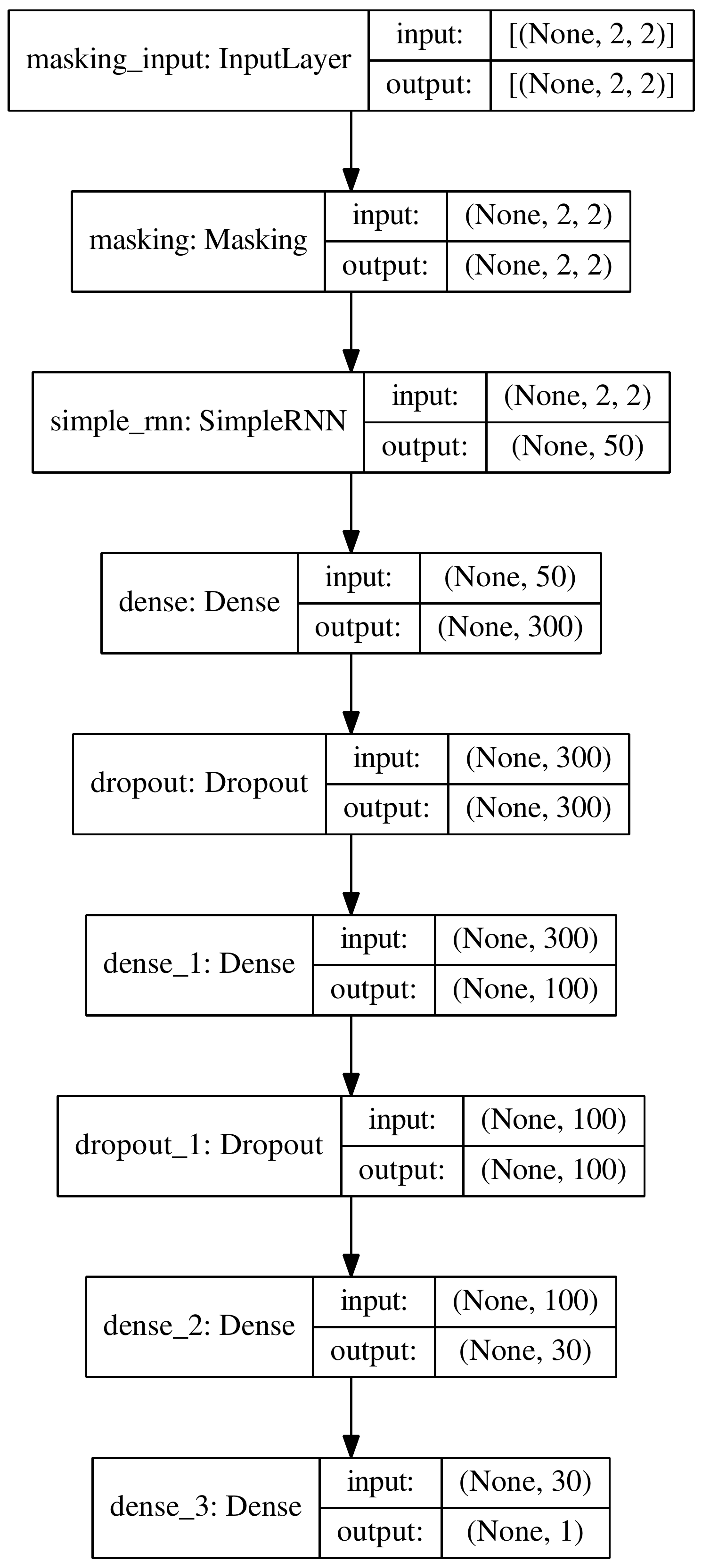}
	\caption{Performance of NMRClass tested on Data B}
\end{figure}

\section{Conclusions}

The separation and purification procedure in organic synthesis is time-consuming. If one can directly ascertain whether the mixture of the organic reaction contains the desired compound structure, it will greatly accelerate 
the rate of total synthesis. In this paper, we attempt to perform some basic exploration  study by using deep learning technology to identify the functional groups or molecular fragments of compounds from the raw NMR data. Some effective techniques are developed to  sample data and treat the problem of data imbalance. Then, based on the structure of RNN, we build a model NMRClass to recognize the functional groups or fragments  of compounds. Compared with the popular models SVM and KNN, the striking benefits are observed.

Although the proposed model NMRClass has achieved good performance in the task of NMR spectrum recognition, we hope the model can make more fundamental contributions to the pattern  recognition of the compound structure in chemistry. If more label association methods are considered, the performance of the model will be further improved. The current study for the functional groups recognition is a key step to the ultimate goal of compounds recognition. In view of this, the following aspects can be discussed in the future:

\begin{itemize}
	\item[1] Collecting huge NMR data and exploring more effective data features. 
	\item[2]  Building more stable and higher performance model.
\end{itemize}

\section*{Acknowledgements}

This work was supported by the National Natural Science Foundation of China under Grant No. 12071195, and the AI and Big Data Funds under Grant No. 2019620005000775.


\end{document}